\documentclass[conference]{IEEEtran}
\usepackage{amsmath,amsfonts,graphicx,mathtools,algorithm,algorithmic,subfigure,multirow,cite,hyperref,float,textcomp}
\usepackage[OT1]{fontenc} 

\DeclareMathOperator{\diag}{\text{diag}}

\title{Fast Color-guided Depth Denoising for RGB-D Images by Graph Filtering}
\begin{document}

\author{\IEEEauthorblockN{Qiwei Huang\IEEEauthorrefmark{1}, Ruikang Li\IEEEauthorrefmark{1}, Zidong Jiang\IEEEauthorrefmark{2}, Wei Feng\IEEEauthorrefmark{2}, Sijie Lin\IEEEauthorrefmark{1}, Hui Feng\IEEEauthorrefmark{1}, Bo Hu\IEEEauthorrefmark{1}}
\IEEEauthorblockA{
\IEEEauthorrefmark{1}Research Center of Smart Networks and Systems, Fudan University, Shanghai, China\\
\IEEEauthorrefmark{2}Beijing iQIYI Science \& Technology Co., Ltd., China\\
\IEEEauthorrefmark{1}\{qwhuang16, rkli17, sjlin18, hfeng, bohu\}@fudan.edu.cn,\IEEEauthorrefmark{2}\{jiangzidong, frankfengw\}@qiyi.com}
}

\maketitle

\begin{abstract}
Depth images captured by off-the-shelf RGB-D cameras suffer from much stronger noise than color images. 
In this paper, we propose a method to denoise the depth images in RGB-D images by color-guided graph filtering. 
Our iterative method contains two components: color-guided similarity graph construction, and graph filtering on the depth signal. Implemented in graph vertex domain, filtering is accelerated as computation only occurs among neighboring vertices. 
Experimental results show that our method outperforms state-of-art depth image denoising methods significantly both on quality and efficiency.
\end{abstract}

\begin{IEEEkeywords}
RGB-D, depth denoising, graph signal processing, graph filtering
\end{IEEEkeywords}

\section{Introduction}
\label{sec:introduction}

RGB-D cameras have been playing important roles in computer vision and robotics \cite{application1, application2, application3} with the ability to obtain 3D geometric information and texture information at the same time, and have even been integrated into high-end cellphones in recent years for applications such as facial recognition and augmented reality (AR), etc..
However, due to the limitation of depth sensors, the quality of the depth images in RGB-D images is usually unsatisfying, suffering from holes, low resolution and especially much stronger noise comparing to that of the color images \cite{rgbd_noise}.

As the depth image and color image both come from one snapshot of the same scene, these two images are spatially corresponded and can be aligned.
In the natural environment, discontinuities often simultaneously present at the same locations in the depth image and its corresponding aligned color image, while regions that are homogeneous in the color image are likely to have similar depth values.
Hence, color images are significant guidelines for the restoration of depth images.
Exploiting the low-noise color images, color-guided methods are effective ways to improve the quality of depth images by fusing depth and color information \cite{AR-model, rcg, color_guided_lowrank_reg, color_guided_filtering, cnndenoise, cnndenoise2}.

Being capable to model the underlying correlations among data, graph signal processing (GSP) \cite{gsp} is an appealing theory to model signals that live on the irregular structure described by graphs \cite{laplacian_matrix}, such as sensor networks and social networks.
A graph can depict the correlations between agents which are modeled as vertices.
The signals of the agents, which are modeled as the graph signal on vertices of the graph, can be further processed with the GSP. 
Recently, GSP theory has been applied to image analysis and processing such as image compression, segmentation, and denoising, etc. \cite{gspcompress, gspcompress2, nlgbt, oglr}.

Color-guided methods for depth image denoising in \cite{color_guided_filtering, rcg, color_guided_lowrank_reg} formulate the problem as an optimization problem and utilize the guidelines from the color images by modifying the regularization.
However, these methods are likely to be intractable and time-consuming to implement.
Another popular framework for color-guided methods is the convolutional neural network (CNN), which denoises the depth images using the trained network, where the color images and original noisy depth images are taken as input \cite{cnndenoise, cnndenoise2}.
Deep learning methods are very computationally intensive and require hardware with high computing capability and a large dataset for training.
Even though GSP has been successfully applied in image denoising \cite{nlgbt, oglr}, the efficiency of graph filtering \cite{denoising_with_graph_filter, denoising_with_trilateral} has not been exploited in the existing methods.
Hence, we firstly introduce graph filtering method to color-guided depth image denoising and make further acceleration.
 
In this paper, we introduce the graph filtering in GSP theory to color-guided depth denoising for RGB-D images.
The proposed iterative method consists of two components: similarity graph construction and graph filtering.
Combining color and depth image data, a similarity graph is constructed, preserving the underlying spatial relationship among pixels.
The depth values of pixels, considered as a graph signal, are denoised with the graph filter designed in graph spectral domain and implemented in vertex domain, which leads to a significant reduction in computational complexity \cite{distributed}.
With a concise framework, our method outperforms state-of-art denoising methods no matter on quality or efficiency.

\section{Preliminaries}
\label{sec:GSP intro}

\subsection{Graph and Graph Signal}
\label{ssec:gsp definition}

An undirected graph without self-loops used to be denoted as $\mathcal{G:=\{V,E},\mathbf{W}\}$, where $\mathcal{V}$, $\mathcal{E}$ and $\mathbf{W}$ are the sets of \emph{vertices}, \emph{edges} and \emph{adjacency matrix}, respectively.
To be more detailed, $v_{i}\in\mathcal{V}$, $(i,j)\in\mathcal{E}$ are considered to be a vertex and an edge connecting vertex $v_{i}$ and $v_{j}$, respectively, subject to $1 \le i, j\le |\mathcal{V}|$.
The \emph{adjacency matrix} is defined as $\mathbf{W}:=[W_{i,j}]_{|\mathcal{V}| \times |\mathcal{V}|}$, where $W_{i,j}$ represents the weight of the edge $(i,j)$, and $W_{i,j}=0$ if edge $(i,j)\notin\mathcal{E}$.
Note that $W_{i,j}\equiv W_{j,i}$ in the undirected graph.
Typically, edge weight $W_{i,j}$ is non-negative, and a large $W_{i,j}$ implies that the samples at vertices $v_i$ and $v_j$ are similar or strongly correlated.

A graph signal refers to the data that resides on the vertices of a graph.
The values sampled at vertex $v_i$ are denoted as $f_i\in\mathbb{R}$.
With the vertices appropriately labeled from 1 to $|\mathcal{V}|$, a graph signal $\mathbf{f}$ on $\mathcal{G}$ can be treated as a vector $\mathbf{f}=[f_1,\dots,f_i,\dots,f_{|\mathcal{V}|}]^\mathrm{T}$, where $\mathbf{f}\in\mathbb{R}^{|\mathcal{V}|}$.

By modeling appropriately, an image can be interpreted as a graph signal on a graph that represents the underlying intrinsic image structure as an image is the set of values on pixels sampled on a 2-D grid.
Generally, an image patch or a pixel is modeled as a vertex in the graph, and edges are connected between vertices following a designed rule \cite{gspimage}.
The intensity on the vertices then is correspondingly considered as the graph signal.

\subsection{Graph Spectrum and Graph Filtering}
\label{ssec:gsp filtering}

Given a graph $\mathcal{G}(\mathcal{V},\mathcal{E},\mathbf{W})$, its \emph{degree matrix} is defined as $\mathbf{D}:=\diag (d_1,\dots,d_i,\dots,d_{|\mathcal{V}|})$, where $d_i = \sum ^{|\mathcal{V}|} _{j=1}W_{i,j}$, and its \emph{Laplacian matrix} is defined as $\mathbf{L}:=\mathbf{D}-\mathbf{W}$\cite{laplacian_matrix}.
As the Laplacian matrix is a positive semidefinite matrix, the eigenvalues of $\mathbf{L}$ are considered as the spectrum of the graph \cite{laplacian1}, \cite{laplacian2}.
With the eigendecomposition, Laplacian matrix can be represented as 
\begin{equation}
\label{equ:decomp}
\mathbf{L=U \Lambda U^{\mathrm{T}}},
\end{equation}
where the eigenvalues $\{\lambda_i\}$ along the diagonal matrix $\mathbf{\Lambda}$ is treated as graph frequencies, and $\mathbf{U}$ is composed of orthogonal eigenvectors $\mathbf{u}_i$ as columns \cite{gspimage}. 
Moreover, the \emph{graph Fourier transform} (GFT) of the graph signal $\mathbf{f}$ is defined as 
\begin{equation}
\label{equ:GFT}
\mathbf{\hat{f}:=U^{\mathrm{T}}f},
\end{equation}
where $\mathbf{\hat{f}}$ is the signal on the graph spectral domain, representing the spectral components of the graph signal $\mathbf{f}$  \cite{laplacian_matrix}.
Accordingly, the inverse GFT is given by $\mathbf{f=U\hat{f}}$.

With the spectrum of a graph signal $\mathbf{f}$, the graph spectral filtering can be defined as $\hat{f}_i^\prime := h(\lambda_i)\hat{f}_i$, where $\hat{f}_i$ is the $i$-th component of $\mathbf{\hat{f}}$, and $h(\cdot)$ is the spectral response function of the filter.
Thus, $\hat{f}_i^\prime$ is the coefficient of frequency component $\mathbf{u}_i$ filtered by $h(\lambda_i)$.
With the GFT and inverse GFT, graph spectral filtering is denoted as 
\begin{equation}
\label{equ:graph filter}
\mathbf{\bar{f}} = \mathbf{U}
\begin{bmatrix}
h(\lambda_1) &  & 0      \\
  & \ddots &  \\
0 &  & h(\lambda_{|\mathcal{V}|})
\end{bmatrix}
\mathbf{U}^{\mathrm{T}}\mathbf{f} = \mathbf{U}h(\mathbf{\Lambda})\mathbf{\hat{f}},
\end{equation}
where $\mathbf{\bar{f}}$ is the filtered graph signal in the vertex domain and $h(\mathbf{\Lambda})$ is the spectral response of the graph filter.

Furthermore, the graph filter can be transformed and fulfilled in the vertex domain, which can be further implemented with an IIR or FIR graph filter \cite{distributed, FIR}.

\section{Proposed Method}
\label{sec:proposed method}

A graph properly generated from a depth image with corresponding aligned color image gives a description of the underlying structure of area segmentation on depth and color jointly.
The depth values on the pixels of the depth image are considered as the graph signal which lives on the graph constructed in the previous step.
With graph spectral analysis, noise-free depth images are manifested to be smooth in graph spectrum.
Treated as a graph signal during the following processing procedure, denoising for depth image is implemented with layers of graph-based filters.

\subsection{Color-guided Similarity Graph Construction}
\label{ssec:depth graph}

\begin{figure}[htbp]
\centering
\includegraphics[width=0.48\textwidth]{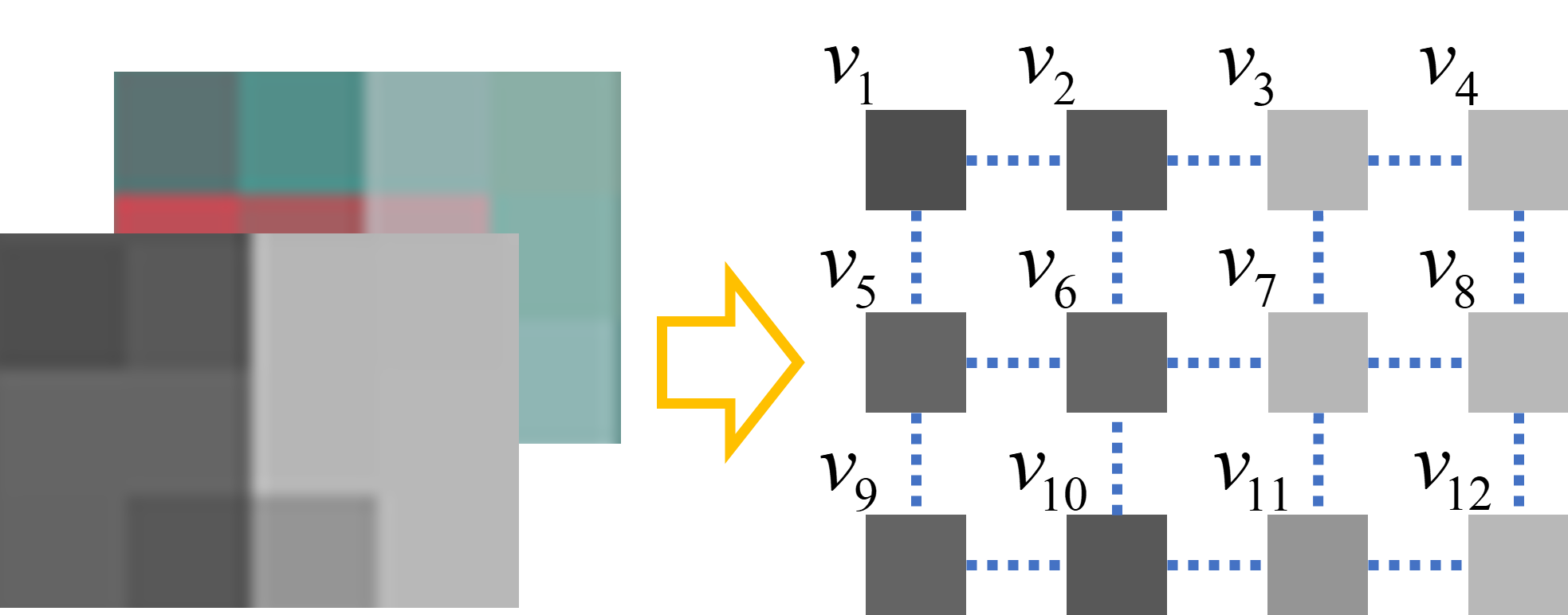}	
\caption{Structure of undirected graph generated from a $4\times3$ RGB-D image.}
\label{fig:depth_graph}
\end{figure}

Consider an $M\!\times\! N$ aligned RGB-D image $\mathcal{P}$, where data on each pixel are denoted as $\mathbf{p}_{m,n}\in\mathcal{P}$, $m\!=\!1,2,\dots,M,n\!=\!1,2,\dots,N$.
To eliminate the influence of lightness, the CIELAB color space is adopted \cite{LAB}.
The data on pixel is defined as $\mathbf{p}_{m,n}\!:=\!(d_{m,n},a_{m,n},b_{m,n})$, where $a_{m,n}$ and $b_{m,n}$ are the values of A and B channel in the CIELAB color space, and $d_{m,n}$ is the depth value.
The topology of similarity graph $\mathcal{G}_\mathrm{depth}(\mathcal{V,E},\mathbf{W})$ is constructed by taking each pixel as a vertex and connecting each pixel to its 4-neighbors to generate the edges, as shown in Fig. \ref{fig:depth_graph} for example.
Designed in left-to-right, top-to-bottom order, each vertex constructed from the pixel is labeled for the modeling of graph signal in the next step.
Depth values of all the pixels are considered as the graph signal which lives on the vertices, which can be treated as a vector $\mathbf{f}\in\mathbb{R}^{MN}$, e.g. $\mathbf{f}=(d_{1,1},d_{1,2},\dots,d_{4,3})^{\mathrm{T}}$ that lives on the graph constructed in the case of Fig. \ref{fig:depth_graph}.

As mentioned in the previous section, the weight $W_{i,j}$ of edge $(i,j)$ implies the correlation between the $i$-th and $j$-th vertices. 
Usually, weight between two pixels is computed using a Gaussian kernel function $\psi(\cdot)$ in the same way it is done in the bilateral filter \cite{bilateral} and graph-based denoising method \cite{nlgbt}.
As the difference between depth values $|d_i-d_j|$ gets larger, the correlation between pixels $v_i$ and $v_j$ gets weaker and the corresponding $W_{i,j}$ gets smaller.
To preserve sharp edges in depth images, e.g. boundaries between the foreground objects and background, we introduce a cut-off policy that sets the weight to zero if $|d_i-d_j|$ is larger than threshold $\Delta_{\mathrm{Th}}$, representing the existence of a sharp edge.
The cut-off policy significantly eliminates potential influence between patches that are `segmented' by weak weight between vertices from the foreground and background.
Furthermore, as discontinuities and similar textured areas in color images are likely to associate with edges and smooth regions in depth images, respectively, the distance between the color information of the two pixels should also be considered in the weight generation function.
Hence, guided with color information, weight is designed as:
\begin{align}
\label{equ:weight formula}
W_{i,j}&=
\begin{cases}
\psi_{\mathrm{d}}(\Delta_{i,j}^{\mathrm{d}})\psi_{\mathrm{a}}(\Delta_{i,j}^{\mathrm{a}})\psi_{\mathrm{b}}(\Delta_{i,j}^{\mathrm{b}}), & \text{if }\Delta_{i,j}^{\mathrm{d}}<\Delta_{\mathrm{Th}}\\
0, & \text{otherwise}
\end{cases}
\end{align}
where $\psi_\mathrm{d}(x)\!=\!\exp\!\left(-\frac{x^2}{2\sigma_\mathrm{d}^2}\right)$, $\psi_\mathrm{a}(x)\!=\!\exp\!\left(-\frac{x^2}{2\sigma_\mathrm{a}^2}\right)$, $\psi_\mathrm{b}(x)\!=\!\exp\!\left(-\frac{x^2}{2\sigma_\mathrm{b}^2}\right)$ are the kernel functions with the pre-designed parameters $\sigma_\mathrm{d}$, $\sigma_\mathrm{a}$, $\sigma_\mathrm{b}$, and $\Delta_{i,j}^{\mathrm{d}}\!=\!|d_{m_1,n_1}\!-d_{m_2,n_2}|$, $\Delta_{i,j}^{\mathrm{a}}\!=\!|a_{m_1,n_1}\!-a_{m_2,n_2}|$, $\Delta_{i,j}^{\mathrm{b}}\!=\!|b_{m_1,n_1}\!-b_{m_2,n_2}|$ with $i\equiv(n_1-1)M+m_1$, $j\equiv(n_2-1)M+m_2$.
Note that the parameters of kernel functions determine the sensitivity to the input, e.g., $\psi_{\mathrm{d}}(\Delta_{i,j}^{\mathrm{d}})$ gets more sensitive to the difference in depth values $\Delta_{i,j}^{\mathrm{d}}$ when $\sigma_\mathrm{d}$ gets smaller. 

\subsection{Graph Filtering}
\label{ssec:filter design}

\begin{figure}[tbp]
\centering
\subfigure[]{
\centering
\includegraphics[width=0.222\textwidth]{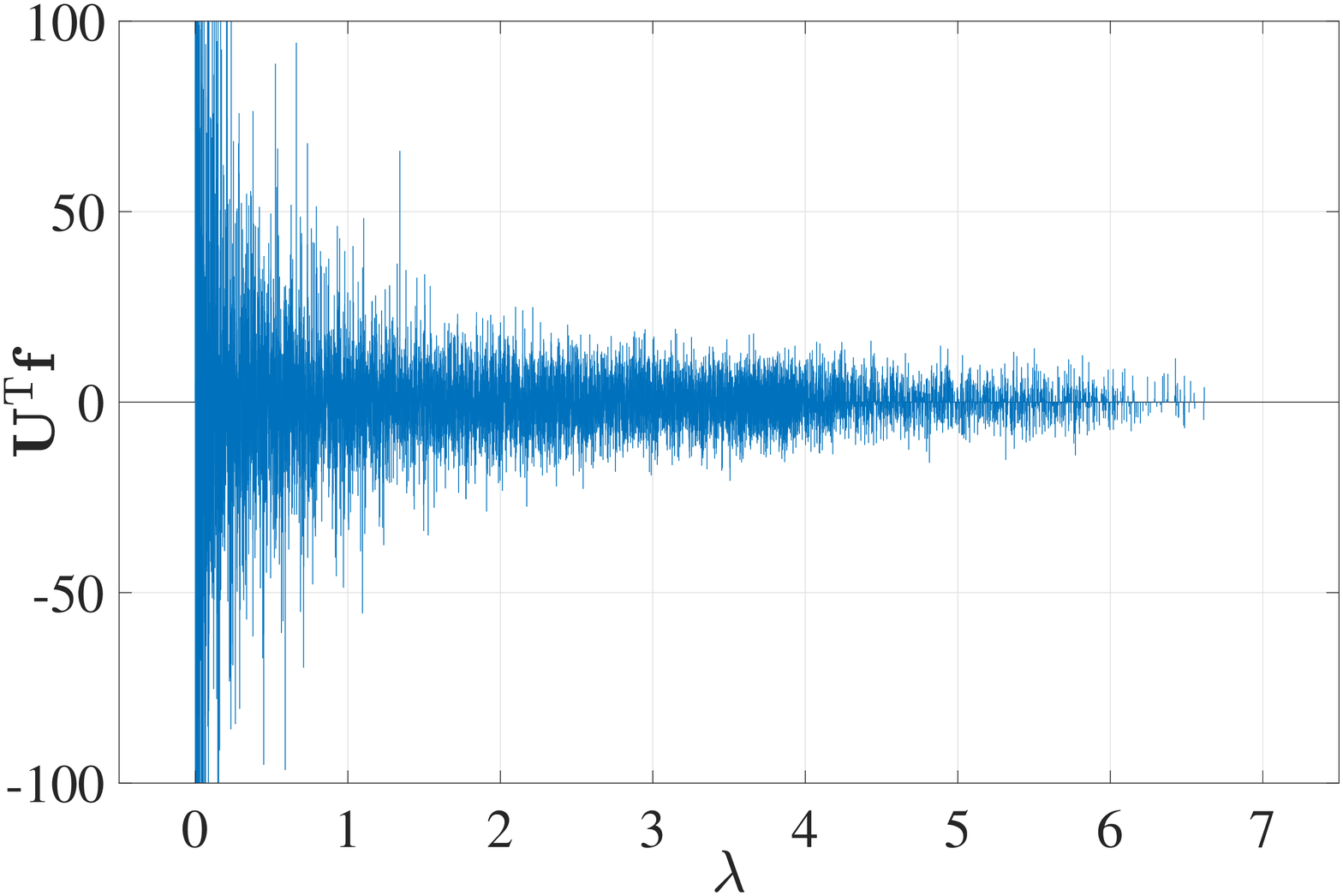}	
}
\subfigure[]{
\centering
\includegraphics[width=0.222\textwidth]{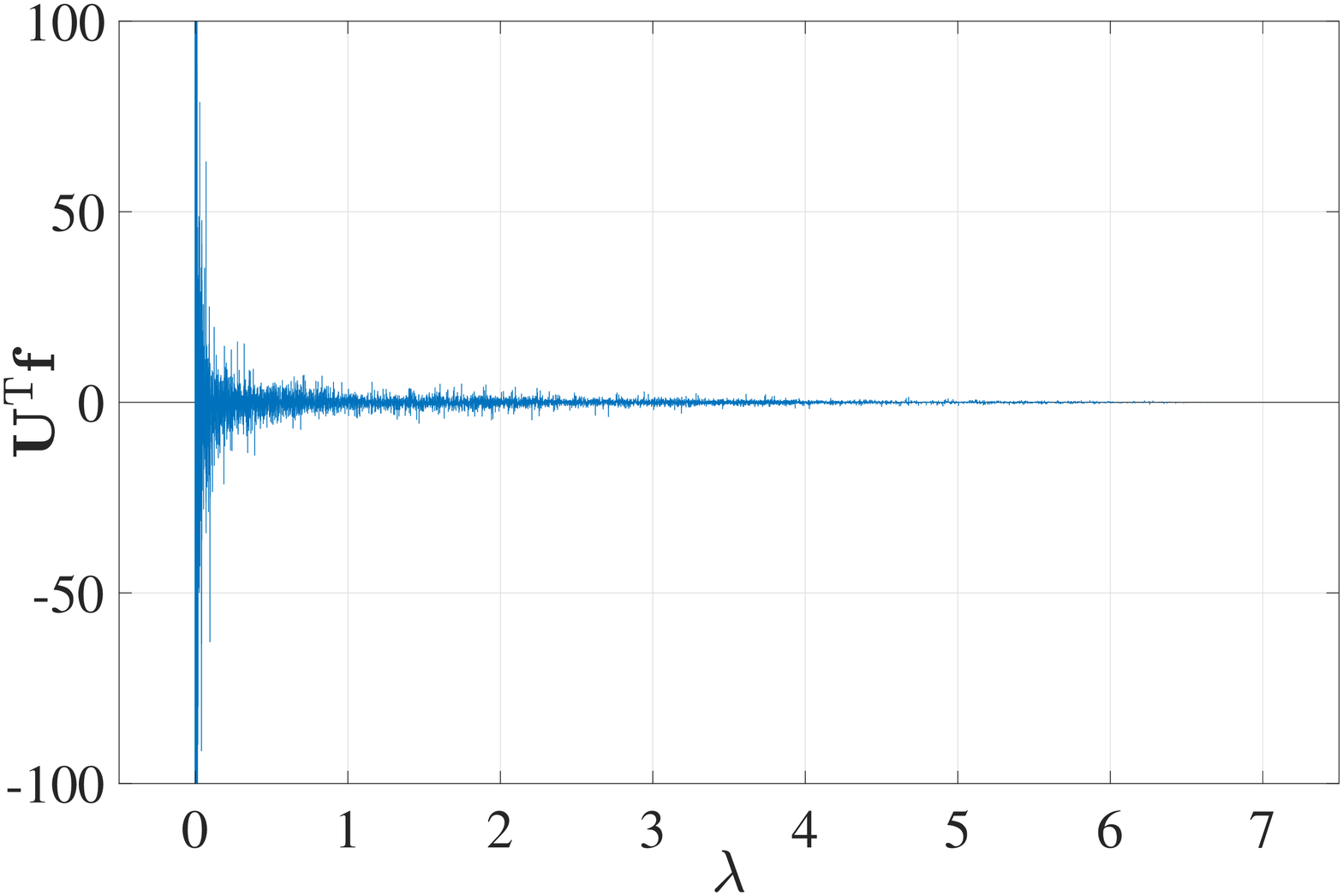}	
}
\caption{Graph spectrum. (a) Noisy depth image. (b) Noise-free depth image.}
\label{fig:spectrum}
\end{figure}

Spectral analysis for the graph signal constructed from the depth data shows that with the piecewise-smooth characteristic \cite{piecewise_smooth}, a noise-free depth image is expected to be dominated by low-frequency components in the graph spectral domain, while a noisy depth image is more likely to get excessive components in the high-frequency region \cite{nlgbt}.
Fig. \ref{fig:spectrum} shows the spectrum of depth signals in the graphs constructed from \textit{Reindeer} in Middlebury stereo datasets \cite{dataset} down-sampled into 112$\times$94, and its noisy version corrupted by additive white Gaussian noise (AWGN) with standard deviation $\sigma=30$.
Therefore, we attenuate the noise with a low-pass graph filter, of which the frequency response function is denoted as $h(\lambda)$ referring to \eqref{equ:graph filter}.
One can customize the low-pass filter and its cut-off frequency appropriately for effective denoising.

After a series of testing for the spectrum of different images in different resolutions and noise levels, it shows that with parameters of \eqref{equ:weight formula} appropriately designed for different noise levels, the spectrum of different noisy depth images in different resolution on the corresponding color-guided similarity graphs is similar in the distribution of power.
Thus, the same graph filter can be deployed for weakening components in the high-frequency region to implement the denoising.

For a small image, the depth image can be denoised in the graph spectral domain using \eqref{equ:graph filter} with ease.
As the image's size expands, the computational complexity in the eigendecomposition of the Laplacian matrix grows explosively.
To be detailed, with the resolution of image expanded to $x$ times, the number of pixels in the image increases to $x^2$ times, such that the size of vertices and Laplacian matrix for the corresponding depth graph increase to $x^2$ times as well.
While the computational complexity of eigenvalue decomposition for $n\times n$ matrix is $O(n^3)$, it is easy to get explosive growth on time consumption when denoising a large image in graph spectral domain, where GFT and inverse GFT for the graph signal generated from the depth image are necessary, leading to a result that eigenvalue decomposition for Laplacian matrix cannot be skipped, referring to \eqref{equ:decomp}, \eqref{equ:GFT}.
Thus, the spectral filter is further implemented in the graph vertex domain to enhance efficiency within the method of graph FIR filter \cite{distributed}.

Specially, further acceleration is made by turning the polynomial function of Laplacian matrix in graph FIR filter into a nested iterative function specified as follow:
\begin{align}
\label{equ:native_FIR}
\mathbf{\bar{f}} & = h(\mathbf{L})\mathbf{f} = (c_0\mathbf{I}+c_1\mathbf{L}+\dots+c_K\mathbf{L}^K)\mathbf{f}\\
\label{equ:FIR}
& = \mathbf{L}\!\left(\cdots\!\left(\mathbf{L}\left(\mathbf{L}\left(c_{K}\mathbf{f}\right)\!+\!c_{K-1}\mathbf{f}\right)\!+\!c_{K-2} \mathbf{f}\right)\!\cdots\!+\!c_{1}\mathbf{f}\right)\!+\!c_{0} \mathbf{f}
\end{align}
where $\mathcal{C}:=\{c_k\},k=0,1,\dots,K$ is the set of coefficients of the $K$-order FIR filter which is accessible by polynomial fitting for the curve of spectral response $h(\lambda)$ about frequency $\lambda\in[0, \lambda_{\mathrm{max}}]$.
Suppose $\mathbf{L}\in \mathbb{R}^{N\times N}$.
The native $K$-order graph FIR filter \eqref{equ:native_FIR} requires $\frac{K(K-1)}{2}N^3+(K+2)N^2$ multiplications with $\frac{K(K-1)}{2}N^3+\frac{K(1-K)+4}{2}N^2-N$ additions, while \eqref{equ:FIR} only requires $KN^2+(K+1)N$ multiplications with $KN^2$ additions.
Hence, \eqref{equ:graph filter} is turned into \eqref{equ:FIR}, in other words, eigenvalue decomposition is no more needed and is replaced by an iterative multiplication of a matrix and a vector, which results in a significant advance in computational efficiency improvement.
Note that distributed computation is available for \eqref{equ:FIR} in graph vertex domain since computation only occurs among neighboring vertices \cite{distributed}.

\subsection{Iterative Denoising with Adaptive Graph Construction}
\label{ssec:FIR filter}
\begin{figure}[bp]
\centering
\subfigure[]{
\centering
\includegraphics[width=0.225\textwidth]{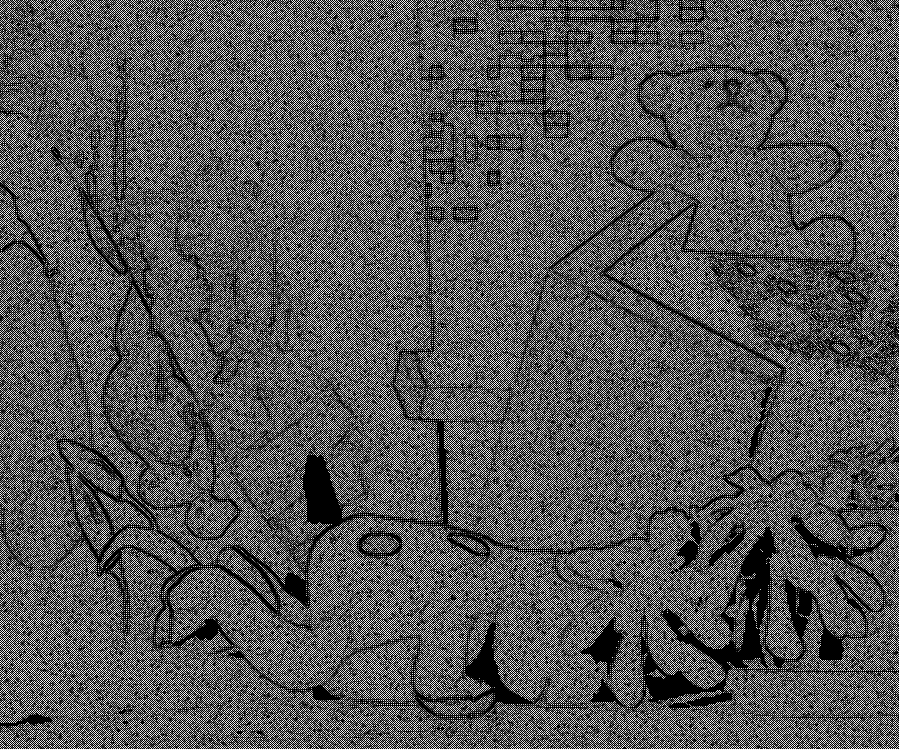}	
}
\subfigure[]{
\centering
\includegraphics[width=0.225\textwidth]{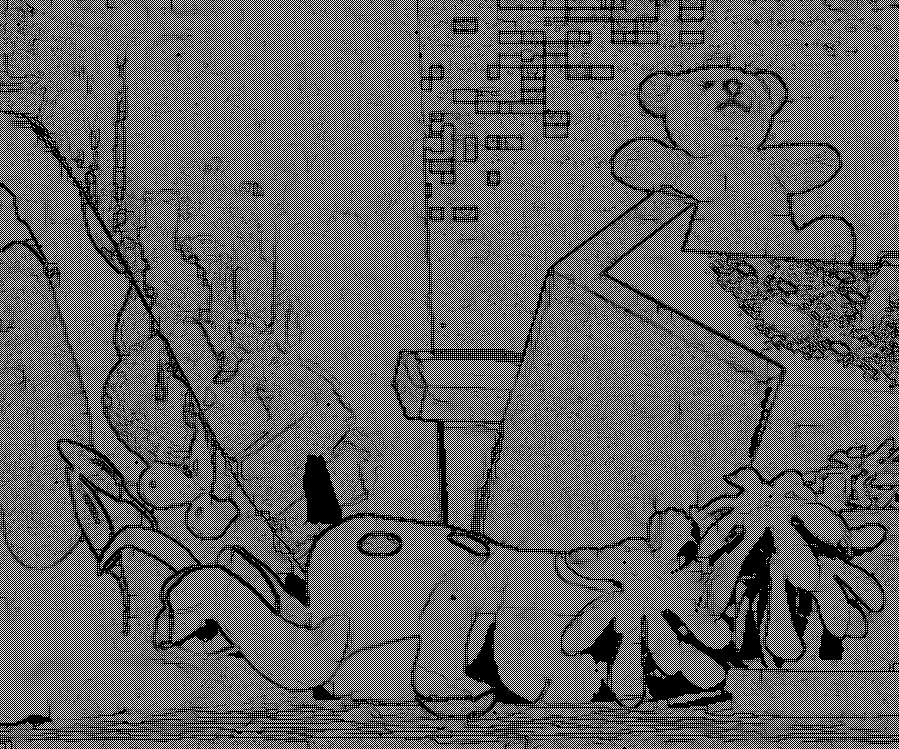}	
}
\caption{Visualized edges of similarity graph of \textit{Teddy} corrupted by AWGN with $\sigma$ = 40. (a) 1st iteration. (b) 6th iteration.}
\label{fig:edges}
\end{figure}
\begin{figure*}[tbp]
\centering
\subfigure[]{
\label{fig:subfig:color}
\includegraphics[width=0.135\textwidth]{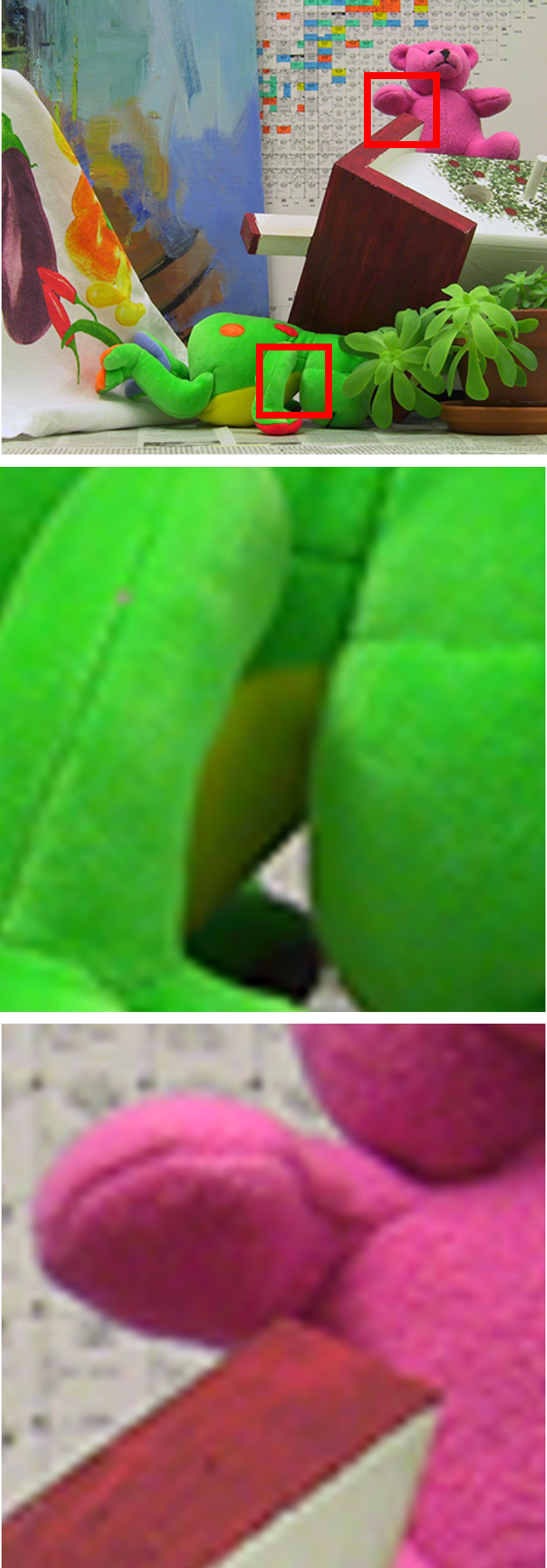}}%
\subfigure[]{
\label{fig:subfig:origin}
\includegraphics[width=0.135\textwidth]{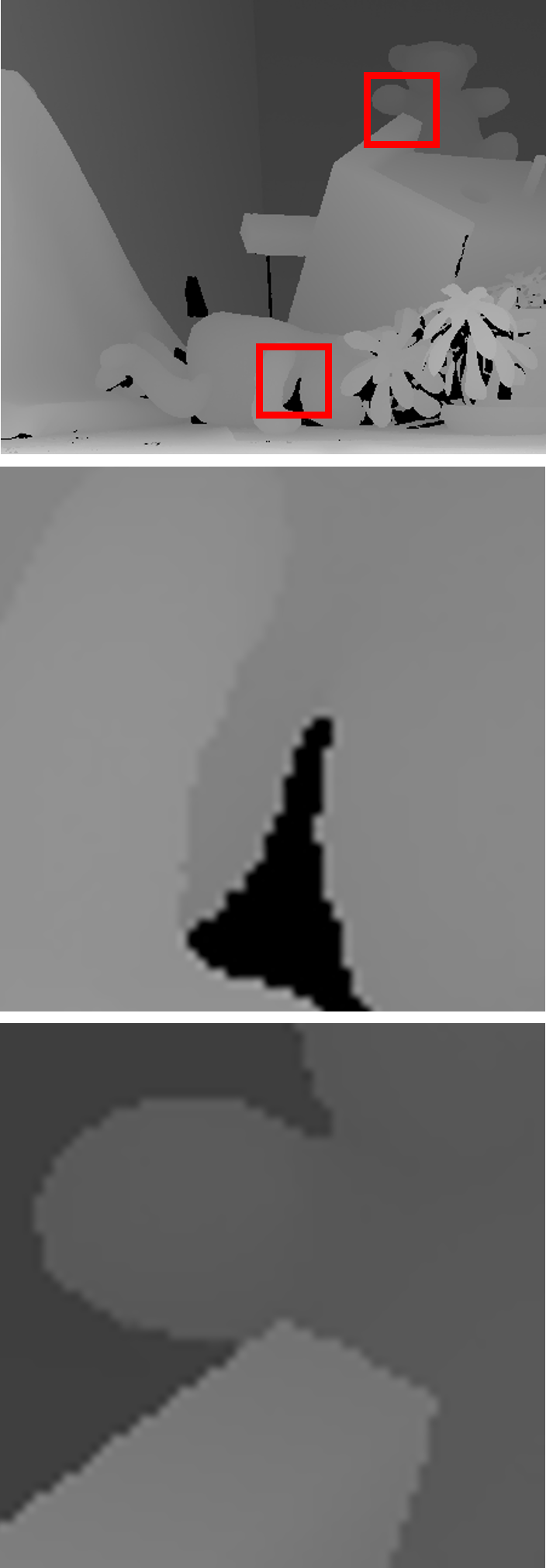}}%
\subfigure[]{
\label{fig:subfig:noisy} 
\includegraphics[width=0.135\textwidth]{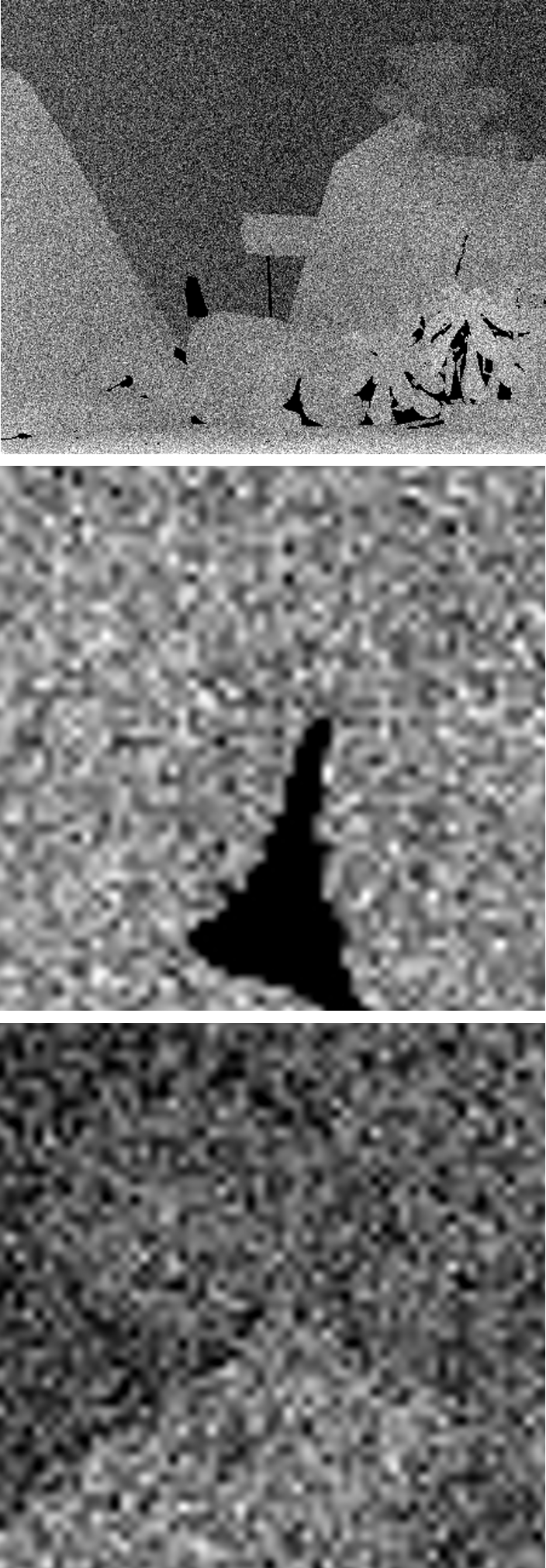}}%
\subfigure[]{
\label{fig:subfig:BM3D} 
\includegraphics[width=0.135\textwidth]{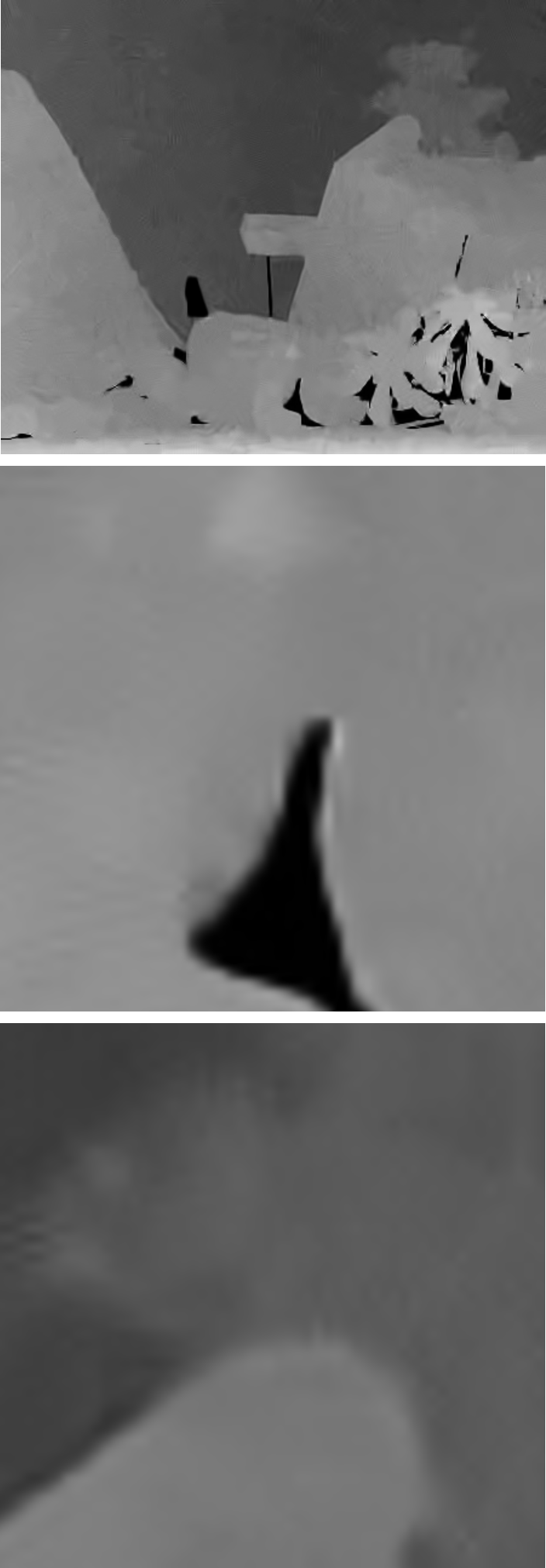}}%
\subfigure[]{
\label{fig:subfig:NLGBT} 
\includegraphics[width=0.135\textwidth]{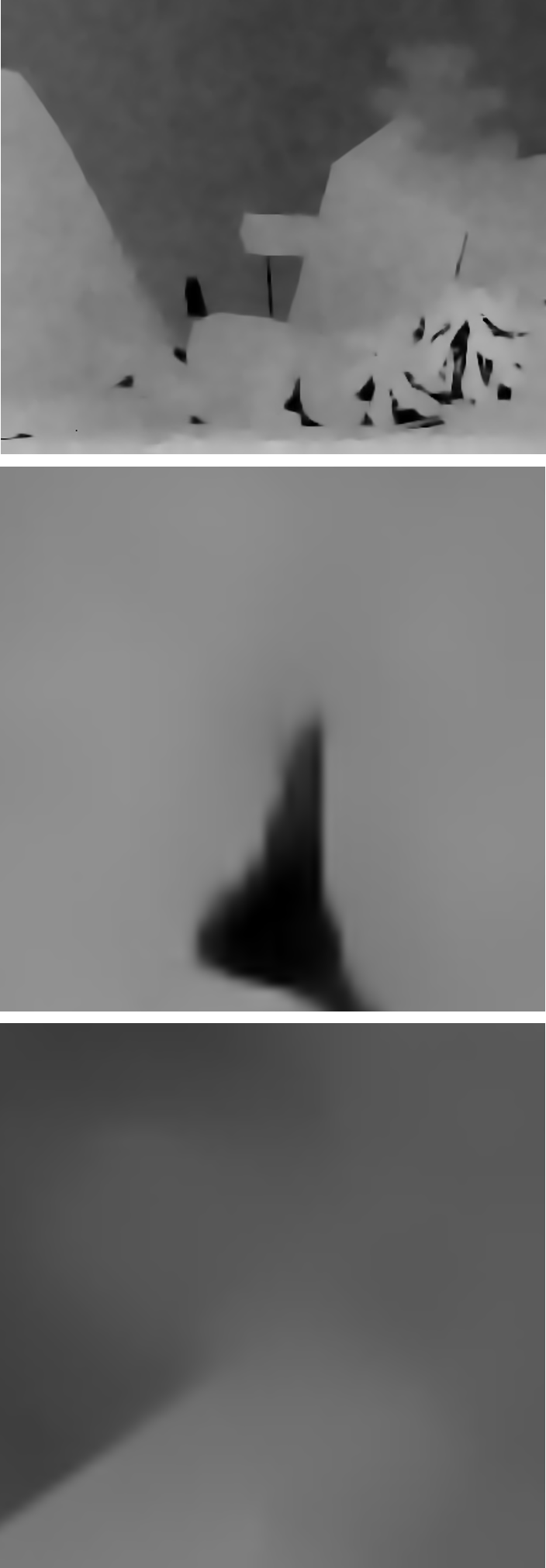}}%
\subfigure[]{
\label{fig:subfig:OGLR} 
\includegraphics[width=0.135\textwidth]{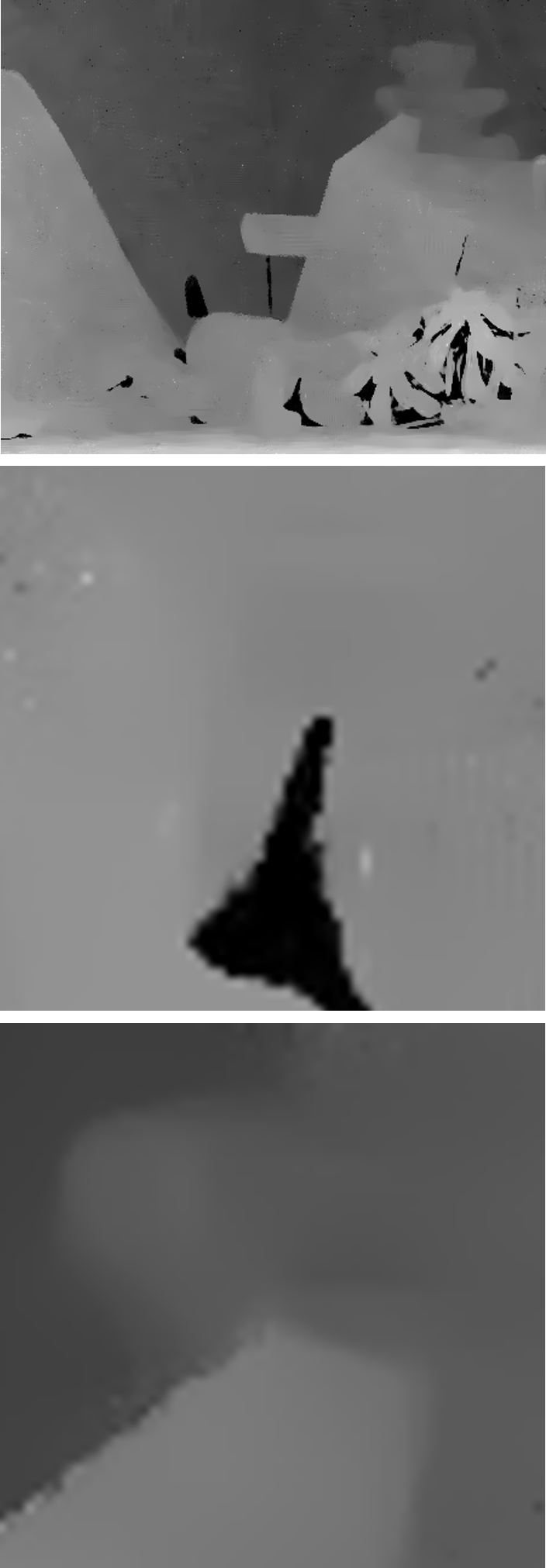}}%
\subfigure[]{
\label{fig:subfig:Ours} 
\includegraphics[width=0.135\textwidth]{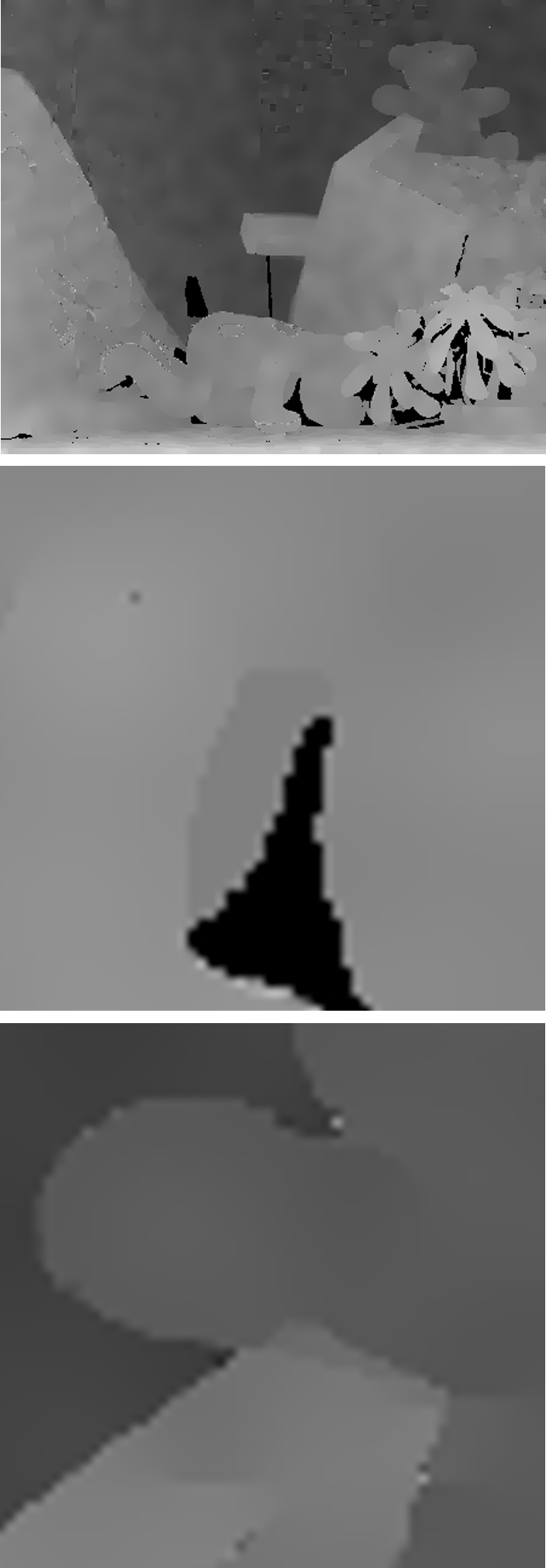}}%
\caption{Denoising of \textit{Teddy} corrupted by AWGN with $\sigma$ = 40. Regions in red boxes are
highlighted for local detail comparison. (a) Color. (b) Noise-free. (c) Noisy 16.38dB. (d) BM3D 31.57dB (e) NLGBT 29.48dB. (f) OGLR 32.25dB. (g) Ours 34.08dB. }
\label{fig:comparision}
\end{figure*}
The denoising in the previous subsection could be improved by repeating it a few more times, which is known as the iterative denoising.
Since the topology of the similarity graph and the graph signal both depend on the depth image, the graph and graph signal should be updated accordingly after the image is filtered.
As the quality of graph topology and signal is improved after each iteration, filtering should become more intensive with different parameters set in the next denoising procedure.
Thus, parameter adaptive tuning for adaptive graph construction is also considered in our method by adding reduction factors $\gamma_{\mathrm{Th}},\gamma_{\mathrm{d}}\in(0,1)$ to $\Delta_{\mathrm{Th}}$ and $\sigma_\mathrm{d}$.
Iterations of denoising with adaptive graph construction are deployed, where rough filtering is assigned in shallow layers to improve the quality of graph topology and signal, while intensive and specific filtering is assigned in deep layers for polishing. 

As shown in Fig.~\ref{fig:edges}, the visualized weighted edges, where the more black indicates the smaller weight, change according to the update of graph signal and topology of similarity graph and become more `strict' but `clean' comparing to the former one in order to eliminate the noise and keep the details for the depth image.
The whole algorithm is described as Algorithm~\ref{alg:Framework}.

\begin{algorithm}[ht] 
\caption{Fast Color-guided Depth Denoising by Graph Filtering.} 
\label{alg:Framework} 
\begin{algorithmic}[1]
	\REQUIRE Noisy RGB-D image $\mathcal{P}$, coefficients of FIR filter $\mathcal{C}$
	\ENSURE Denoised image $\bar{\mathcal{P}}$
	\STATE (Initialization) obtain $M$, $N$, set $\bar{\mathcal{P}}^{(0)}=\mathcal{P}$, set parameters $\Delta_{\mathrm{Th}}^{(1)}$, $\sigma_\mathrm{d}^{(1)}$, $\sigma_\mathrm{a}$ and $\sigma_\mathrm{b}$, set reduction factors $\gamma_{\mathrm{Th}}$ and $\gamma_{\mathrm{d}}$
	\FOR {$t = 1$ to $T$}		
		\label{code:fram:param}
		\STATE $\mathcal{P}^{(t)}=\bar{\mathcal{P}}^{(t-1)}$, construct $\mathcal{G}_\mathrm{depth}^{(t)}$ and $\mathbf{f}^{(t)}$ by \eqref{equ:weight formula}
		\label{code:fram:construct}
		\STATE Compute $\mathbf{\bar{f}}^{(t)}$ by \eqref{equ:FIR}
		\STATE Reconstruct denoised image $\bar{\mathcal{P}}^{(t)}$ from $\mathbf{\bar{f}}^{(t)}$, update $\Delta_{\mathrm{Th}}^{(t+1)}=\gamma_{\mathrm{Th}}\Delta_{\mathrm{Th}}^{(t)}$, $\sigma_\mathrm{d}^{(t+1)}=\gamma_{\mathrm{d}}\sigma_\mathrm{d}^{(t)}$
	\ENDFOR
\RETURN Denoised image $\bar{\mathcal{P}}=\bar{\mathcal{P}}^{(T)}$
\end{algorithmic}
\end{algorithm}

\section{Experiment}
\label{sec:experiment}

\subsection{Experimental Setup}
\label{ssec:experiment setup}

We evaluate our method with different sizes of depth images together with their aligned color images: \textit{Teddy, Moebius}, and \textit{Dolls} in Middlebury stereo datasets \cite{dataset}.
Additive white Gaussian noise (AWGN) with standard deviation $\sigma$ ranging from 10 to 50 is added to the depth images as the input noisy depth images. 
A 2-order Butterworth filter is deployed for the graph filtering with cut-off frequency empirically designed as $\lambda_{\mathrm{max}}/43$.
In general, 6-10 iterations are deployed for iterative denoising.
Implemented in MATLAB\textregistered\ R2018b on a desktop with an Intel\textregistered\ Core\texttrademark\ i7-4790K CPU, performance comparisons in average PSNR and average time consumption  with BM3D \cite{bm3d}, NLGBT \cite{nlgbt} and OGLR \cite{oglr} are made.

\subsection{Experimental Results}
\label{ssec:results}
\begin{table}[tbp]
\centering
\caption{Performance comparison in average PSNR (dB)}
\label{tbl:psnr}
\begin{tabular}{|c|c|c|c|c|c|c|}
\hline
Image                     & Method & $\sigma$=10    & $\sigma$=20    & $\sigma$=30    & $\sigma$=40    & $\sigma$=50 \\ \hline
\multirow{4}{*}{Teddy}    & BM3D   & 41.17          & 35.94          & 33.16          & 31.32          & 29.73       \\
                          & NLGBT  & 41.80          & 36.84          & 33.85          & 31.65          & 30.26 \\
                          & OGLR   & 42.80          & 37.73          & 34.52          & 32.20          & 30.70 \\
                          & Ours   & \textbf{43.64} & \textbf{38.80} & \textbf{36.12} & \textbf{34.11} & \textbf{32.96} \\ \hline
\multirow{4}{*}{Moebius}  & BM3D   & 42.03          & 37.15          & 34.70          & 33.09          & 31.75 \\
                          & NLGBT  & 42.58          & 37.63          & 34.89          & 33.13          & 31.98 \\
                          & OGLR   & 43.31          & 38.36          & 35.35          & 33.19          & 31.94 \\
                          & Ours   & \textbf{43.73} & \textbf{39.55} & \textbf{37.15} & \textbf{35.81} & \textbf{34.53} \\ \hline
\multirow{4}{*}{Dolls}    & BM3D   & 40.77          & 35.91          & 33.56          & 32.19          & 30.87 \\
                          & NLGBT  & 41.75          & 37.21          & 33.96          & 31.63          & 30.44 \\
                          & OGLR   & 42.54          & 37.87          & 34.91          & 32.63          & 31.43 \\
                          & Ours   & \textbf{43.03} & \textbf{39.26} & \textbf{37.11} & \textbf{35.60} & \textbf{34.34} \\ \hline
\end{tabular}
\end{table}
\begin{table}[tbp]
\centering
\caption{Performance comparison in average time consumption (sec)}
\label{tbl:efficiency}
\begin{tabular}{|c|c|c|c|}
\hline
Method & 463$\times$370	& 695$\times$555 & 1390$\times$1110 \\ \hline
BM3D   & 2.1            & 5.5            & 22.1             \\ \hline
NLGBT  & 47.8           & 112.1          & $>$300           \\ \hline
OGLR   & 184            & $>$300         & $>$300           \\ \hline
Ours   & \textbf{1.8}   & \textbf{3.8}   & \textbf{14.8}    \\ \hline
\end{tabular}
\end{table}
\textbf{Objective results.}
The performance comparisons with different methods are shown in Table~\ref{tbl:psnr} and Table~\ref{tbl:efficiency}, where the best one is marked in bold in each comparison.
For the performance of denoising, PSNR is taken as the quantitative indicator by calculating the difference between the denoised image and the noise-free image.
As shown in Table~\ref{tbl:psnr}, the proposed method turns out to outperform other methods more significantly when dealing with noisier depth images.
Additionally, time consumption comparisons show that the proposed method is much more efficient than other methods.

\textbf{Subjective results.}
As illustrated in Fig.~\ref{fig:comparision}, the denoising results from \textit{Teddy} corrupted by AWGN with $\sigma=40$ are listed.
The proposed method significantly outperforms the competing methods, especially in the restoration and preservation of the sharp edge between the foreground and background.
With the guidelines from the low-noise color image, the restoration for edges corrupted totally by the noise is easily and properly fulfilled using the proposed method.

\section{Conclusion}
\label{sec:conclusion}
In this paper, we newly introduce the graph filtering to the color-guided depth denoising for RGB-D images.
Experimental results show that the proposed method, the fast color-guided depth denoising by graph filtering, is effective, efficient and easy to implement.
Noticing that our iterative denoising methods with parameters preset is similar to the method of graph convolutional neural network (GCNN), which treat the graph construction and filter designing in the same way the deep learning does, we would like to solve the problem with GCNN and compare with the method proposed here, and look forward to discover the relations between them in the future.

\section*{Acknowledgement}
\addcontentsline{toc}{section}{Acknowledgement} 
This work was supported by Beijing iQIYI Science \& Technology Co., Ltd. and the National Natural Science Foundation of China (NSFC) under Grant No. 61501124.

\bibliographystyle{IEEEbib}
\bibliography{Reference}

\end{document}